\documentclass[fleqn,10pt]{wlscirep}

\usepackage[utf8]{inputenc}
\usepackage[T1]{fontenc}

\usepackage{amsmath,amssymb,amsfonts}
\usepackage{graphicx}
\usepackage{float}
\usepackage{textcomp}
\usepackage{xcolor}
\usepackage{graphicx}%
\usepackage{multirow}%
\usepackage{lipsum}
\usepackage{amsmath,amssymb,amsfonts}%
\usepackage{amsthm}%
\usepackage{mathrsfs}
\usepackage{wrapfig}%
\usepackage{xcolor}%
\usepackage{braket}
\usepackage{textcomp}%
\usepackage{manyfoot}%
\usepackage{listings}%
\usepackage{xcolor}
\usepackage{soul}
\usepackage{booktabs}%
\usepackage{algorithm}%
\usepackage{algorithmicx}%
\usepackage{algpseudocode}
\usepackage{enumitem}

\title{An Improved Quantum Anonymous Notification Protocol for Quantum-Augmented Networks}

\author[1,*]{Nitin Jha}
\author[1, +]{Abhishek Parakh}
\author[2, +]{Mahadevan Subramaniam}
\affil[1]{Kennesaw State University, GA, USA}
\affil[2]{University of Nebraska Omaha, NE, USA }

\affil[*]{Corresponding author: njha1@students.kennesaw.edu}
\affil[+]{This work is partly sponsored by NSF award \#2324924 and \#2324925.}

\begin{abstract}
The scalability of current quantum networks is limited due to noisy quantum components and high implementation costs, thereby limiting the security advantages that quantum networks provide over their classical counterparts. Quantum Augmented Networks (QuANets) address this by integrating quantum components in classical network infrastructure to improve robustness and end-to-end security. To enable such integration, Quantum Anonymous Notification (QAN) is a method to anonymously inform a receiver of an incoming quantum communication. Therefore, several quantum primitives will serve as core tools, namely, quantum voting, quantum anonymous protocols, quantum secret sharing, etc. However, all current quantum protocols can be compromised in the presence of several common channel noises. In this work, we propose an improved quantum anonymous notification (QAN) protocol that utilizes rotation operations on shared GHZ states to produce an anonymous notification in an n-user quantum-augmented network. We study the behavior of this modified QAN protocol under the dephasing noise model and observe stronger resilience to false notifications than earlier QAN approaches. The QAN framework is also proposed to be integrated with a machine-learning classifier, an enhanced quantum-augmented network. Finally, we discuss how this notification layer integrates with QuANets so that receivers can allow switch-bypass handling of quantum payloads, reducing header-based information leakage and vulnerability to targeted interference at compromised switches.
\end{abstract}
\begin{document}

\flushbottom
\maketitle
%
%
\thispagestyle{empty}

\noindent Please note: Abbreviations should be introduced at the first mention in the main text – no abbreviations lists. Suggested structure of the main text (not enforced) is provided below.

\section{Introduction}
Quantum Communication is one of the key aspects of the quantum internet, similar to how communication is an integral part of the current age internet. Current age quantum internet research faces several scalability issues similar to early models of the internet. Other than scalability issues, the practical realization of quantum protocols has always faced several hardware issues. All the initial work regarding quantum key distribution (QKD) protocols, such as BB84 \cite{bennett2014quantum}, B92 \cite{bennett1992quantum}, etc., is based on ideal single-photon sources. However, in practice, ideal single-photon sources are not practical. This has been worked out by using other imperfect single-photon sources, such as heralded spontaneous parametric down-conversion (SPDC), to simulate single-photon sources.  \cite{yuan2016simulating} Several works pointed out the usefulness of the multiphoton approach as a possible solution to several challenges associated with single photon sources.  \cite{burr2022evaluating, pont2023multi, Jha2025MultiPhotonQKD} Multi-photon QKD protocols have been studied as a possible solution to several noises, such as decoherence, bit-flip, phase-flip, etc.  \cite{jha2024effect} However, there is a problem associated with the multiphoton approach, namely higher risk of siphoning attacks, lesser key-rates due to increased occupancy of bandwidth (same number of bits have more corresponding qubits), etc \cite{jha2025quantumreview}. Therefore, we need to study other possible solutions to noise in quantum devices that can address both the security and scalability issues; one such approach can be designing quantum communication protocols that have inbuilt error correction codes \cite{jha2024joint}.

\begin{figure*}[h!]
    \centering
    \includegraphics[width=0.8\linewidth]{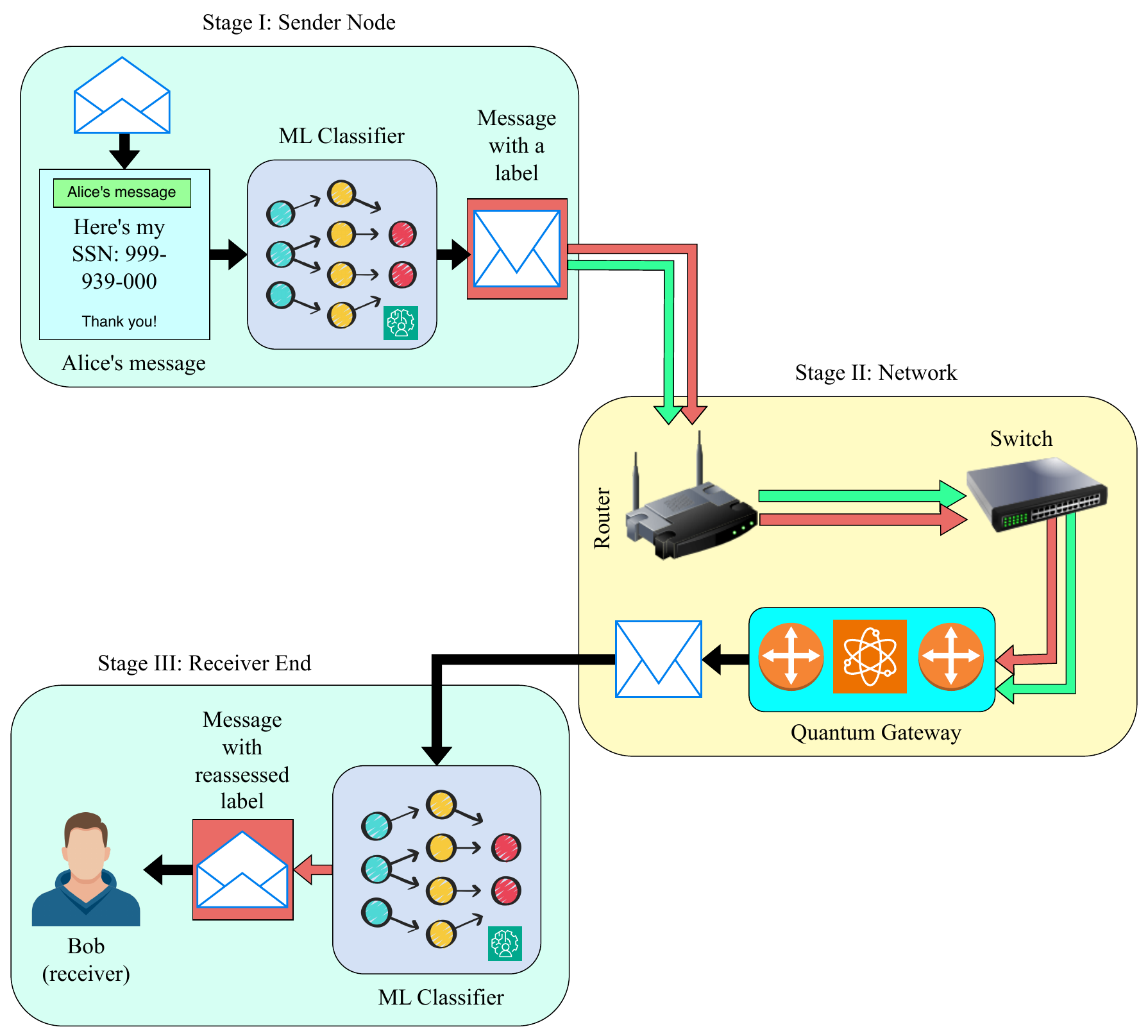}
    \caption{One vision of quantum augmented network, as presented in  \cite{jha2024ml}. The work presents the idea of using machine learning algorithms to selectively classify communication messages in ``private'' and ``non-private'' classes. This allows us to only use quantum encryption for private classes and save resources for non-private messages. In the diagram, the \textbf{red} arrow indicates the flow of private/quantum-encryption information, and the \textbf{green} arrow indicates the flow of non-private/classical encrypted information. }
    \label{fig:quanet}
\end{figure*}

Quantum Augmented Networks (QuANets) is a recent development leading to a scalable and practical idea of a global quantum internet by combining several quantum primitives such as Quantum Anonymous Notification (QAN) protocols, Quantum Key Distribution, Quantum Secure Direct Communication, etc, into several currently existing classical infrastructure \cite{pan2024evolution}. As the quantum components remain ridden by several noises such as \textit{decoherence}, \textit{attenuation}, etc., the scalability of purely quantum networks is severely limited. These noises reduce the fidelity of the quantum states over large distances.  The preliminary model of QuANet used in this paper was proposed by the authors in  \cite{jha2024ml}. Figure. (\ref{fig:quanet}) describes the working of this model. It consists of three stages as explained in  \cite{jha2025towards}: (1) ML classifier assigns a privacy label to each message, determining if there's any private content, (2) the message is selectively encrypted using quantum encryption if it has any private content, (3) transmitting the message packet, and then reaching their destinations. The outgoing packet in this method consists of the following parts: (1) Headers: considers information regarding classical or quantum encryption, addresses, etc., and (2) Payloads: quantum-encrypted or classical-encrypted payload. There are certain core vulnerabilities of this approach, especially in the case of networks having compromised switches.

A critical vulnerability in this QuANet arises from the possibility of compromised switches. In a practical network, we have to assume the presence of several compromised switches and limited trusted ones. Considering the above model of QuANet, we consider the class of attacks performed by compromised switches. For example, if we include information about a packet header containing quantum payload, then an attacker can use this information to flag packets having sensitive information, leading to traffic inference attacks. An adversary can selectively attack packets that have a quantum payload by either delaying or dropping them.  There can also be several attacks based on network monitoring rather than active attacks. Therefore, the practical realization of QuANets depends not only on the robust quantum encryption techniques but also on mitigating several network-level attacks as mentioned above.

\subsection{Related Works}
Based on the above-mentioned issues, we introduce the use of the quantum anonymous notification (QAN) protocol as an extra layer in the quantum augmented network framework. A general QAN protocol allows a user in the network to anonymously notify any other user in the network using pairs of pre-shared GHZ states \cite{khan2020quantum}. {Several earlier works explore anonymous primitives for quantum networks from both theoretical and experimental perspectives, including notification, collision detection, and anonymous message transmission} \cite{khan2020quantum, dejong2023anonymous, ruckle2023experimental, grasselli2022secure, yang2021towards, huang2022experimental, unnikrishnan2019anonymity, gong2022anonymous}. 

{The work by Khan et al. proposed one of the popular formalizations of QAN, relying on GHZ correlations to establish sender anonymity while ensuring message delivery }\cite{khan2020quantum}. {However, these early approaches were primarily theoretical and did not account for realistic imperfections in the physical layer, such as photon loss, depolarizing noise, or detector inefficiencies. In realistic optical and fiber-based networks, photon loss, dephasing, depolarizing noise, and detector inefficiencies reduce multipartite entanglement quality, which can directly impact the false-notification probability and, thus complicate anonymity verification at larger network scale. This practical gap motivates follow-on efforts that either (i) revise the entanglement resource, (ii) formalize better and more robust security and correctness notions, or (iii) demonstrate feasibility based on deployed network infrastructure, i.e., analysis under some basic noise models.}

{A major line of work explores anonymous conference key agreement (ACKA) and other related multi-party anonymity primitives as notification-adjacent building blocks. For example, de Jong et al. propose ACKA using linear cluster states constructed over repeater-like chains, emphasizing practical network resource assumptions where nodes only need neighbor entanglement rather than a central GHZ distributor }\cite{dejong2023anonymous}. {R\"uckle et al. demonstrate an experimental demonstration of anonymous conference key agreement (ACKA) using four-photon linear cluster states and explicitly optimize protocol parameters under finite-size and noise considerations} \cite{ruckle2023experimental}. {Grasseli et al. demonstrated that GHZ based protocols can outperform protocols generally based on bipartite entanglement state and provides stronger anonymity guarantees} \cite{grasselli2022secure}. {Another work by Yang et al. introduced anonymous quantum transmission mechanisms that rely on EPR-based entanglement distribution, enabling message transfer without revealing sender or receiver identities} \cite{yang2021towards}.

{Large-scale feasibility has also been demonstrated in city-scale experiments involving up to eight users, showing the potential of GHZ and cluster-state-based architectures for practical anonymous communication by Huang et al.} \cite{huang2022experimental}. {Some other works such as by Unnikrishnan et al. proposed one of the first realistic implementations of anonymous quantum communication under noise, showing how anonymity can be preserved even with imperfect entanglement}  \cite{unnikrishnan2019anonymity}. {Gong et al. presented a W-state-based anonymous communication framework that incorporates both notification and collision detection sub-protocols, directly aligning with the goals of QAN systems} \cite{gong2022anonymous}. {Furthermore, Zheng et al. investigated collision detection in multi-user networks by leveraging higher-partite entangled states, highlighting how quantum resources can enable contention-free medium access in anonymous communication settings}  \cite{zheng2024anonymouscollision}. {Such collision detection and notification framework is quite relevant in terms of QuANet-like deployment where we assume the quantum resources are scarce. In this regime, coordination is first constraint and badly designed control signaling can leak important metadata that can risk the anonymity of the entire network. }

{Despite these advances, existing QAN protocols often overlook the impact of environmental noise and decoherence, which can significantly degrade entanglement fidelity and, consequently, the reliability of anonymity verification. In this work, we propose an improved QAN protocol that uses rotation operations to introduce a phase change that results in a parity flip after measurement by the receiver. This modified QAN protocol shows a lower probability of false notifications in the case of channel noise. }The following are the main contributions of this work:

\begin{enumerate}
    \item {The QAN protocol has been studied under two types of dephasing noise models, while providing anonymity to both the user and the receiver.}

    \item {The integration mechanism with QuANet is presented.}

    \item {This integration addresses a key vulnerability: context-bearing headers that reveal the presence of quantum payloads and enable targeted interference. This integration makes ``switch independence" possible, where the receiver can allow for a switch-bypass in case of an anonymous notification. This reduces contextual leakage at compromised switches, limits possible adversarial traffic selection, and preserves end-point anonymity.}
\end{enumerate}

This paper is structured as follows. Sec \ref{Sec:qan} defines the quantum anonymous protocol and provides a test case showing the working of this protocol. In this section, we also present simulation results for the working of the QAN protocol, a comparison of the accuracy of our improved QAN protocol with older approaches under noise models, and analyze two possible attacks on this system. Sec \ref{Sec:integrate} integrates the idea of quantum augmented networks with the use of quantum anonymous protocol, providing a foundation for future work on integrating these ideas into a full communication stack.

\section{Methods}
\label{Sec:qan}

\subsection{The Quantum Anonymous Notification Protocol}

There are several approaches to quantum networks which are based on quantum voting \cite{vaccaro2007quantum, li2021quantum, arapinis2021definitions}, quantum secret sharing \cite{hillery1999quantum, gottesman2000theory, senthoor2022theory}, quantum conference agreement \cite{murta2020quantum, pickston2023conference}, etc. Early works regarding the quantum internet \cite{kimble2008quantum} were based on the ideas mentioned above. One key aspect of any communication network is preserving the privacy of the involved users. Therefore, even in the case of quantum internet, we need to utilize different privacy-preserving techniques to guarantee private communications. A Quantum Anonymous Notification protocol allows a sender to notify the intended receiver of an incoming communication while preserving the identity of the involved parties. Earlier works on QAN focus mainly on preserving the identity of either the receiver or the sender; however, we propose a modified QAN protocol that preserves the identity of both users.  

\begin{figure*}[h!]
    \centering
    \includegraphics[width=0.6\linewidth]{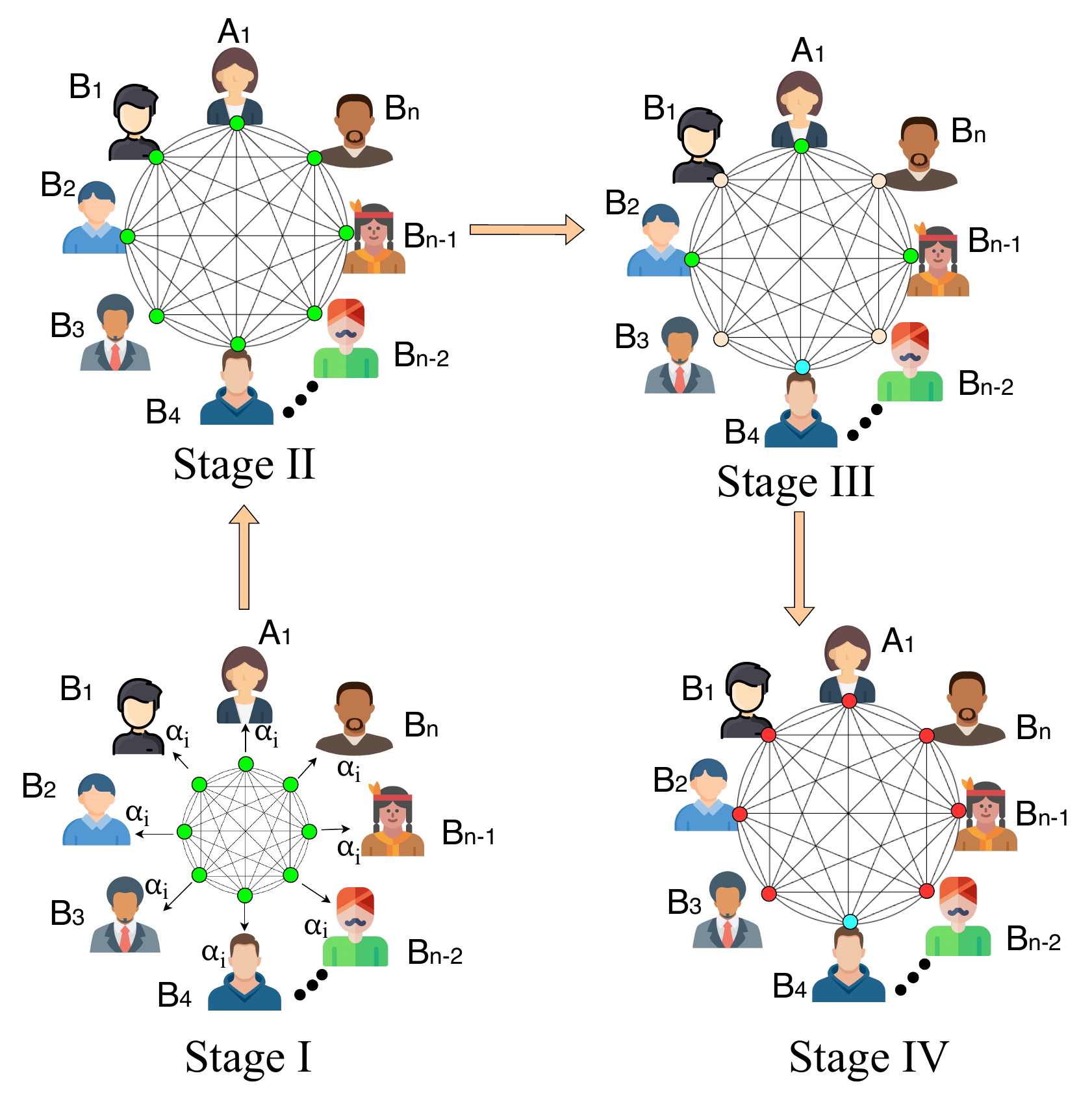}
    \caption{A schematic diagram representing different stages of the outlined QAN protocol. The diagram shows a network having $ n$-users, \textit{A} represents Alice, and all other nodes are marked $B_i\in [1, n]$, as each of them is equally likely to be the receiver of the notification by Alice. \textit{Stage I} is distributing the secret share of the angle $\alpha$, and the GHZ state. \textit{Stage II} represents each node holding a state, and waiting to perform the relevant rotation operation. {Stage I and II consists of Steps 1 and 2 from the detailed protocol description below.} \textit{Stage III} represents the execution of the QAN. Green nodes show the application of $R_z(\alpha')$, and yellow nodes show no rotation application. {Stage III shows a schematic representation of Steps 3 through 5, i.e., till measurement is performed.} \textit{Stage IV} is the final stage of the QAN protocol, where Bob's (receiver, marked by $B_4$) node is highlighted in blue, and every other node is marked by red, {which is the Step 5 of the protocol where only the designated receiver has the notification}. This signifies that Bob identifies a notification upon doing the post-measurement parity calculation. This diagram serves to give the user a visual idea about the working of the protocol, with detailed protocol steps explained subsequently.}
    \label{fig:qan-diag}
\end{figure*}

In an earlier work \cite{khan2020quantum}, the authors propose an anonymous quantum private comparison protocol in an $n$-node quantum network using quantum anonymous notification protocol (QAN). In this work, we modify the QAN protocol such that instead of using two unitary operations, we share parts of an angle ($\alpha$) using secret sharing, and then use this to execute the QAN. {Apart from this, we apply a layered approach where each user have two sets of GHZ states which allows for enhanced anonymity. We also include a study of the working of this modified protocol in presence of depolarizing noise to simulate realistic hardware setups}. The exact details of this modification are as follows, 

\begin{enumerate}[align=left]
    \item[\textbf{Step 1:}] Distributing GHZ states.

    We assume the network constitutes of $n$ users ($1,2 \cdots, n$), and each party distributes one $n-$partite GHZ state among the $n$ parties as follows, 
    \begin{equation}
        |\text{GHZ}\rangle = \frac{1}{\sqrt{2}}\left(\ket{0}^{\otimes n} + \ket{1}^{\otimes n} \right).
        \label{GHZ-EQ}
    \end{equation}
    They then send exactly one qubit to each of the other $(n-1)$ parties, along with the local index.  Concretely, when party $j$ sends a qubit to party $i$, they tagged as $\Phi_j^{-1}(i)$ of GHZ$_j$.”

    Thus recipient $i$ learns both
    \begin{enumerate}
        \item[a.] which GHZ state it came from ($j$), and  
        \item[b.] which position in that state they hold ($\Phi_j^{-1}(i)$), 
    \end{enumerate} 
    without ever learning the full bijection $\Phi_j$.  We denote that qubit by $q_j^i$—the $i^\text{th}$ user’s share of $\ket{\mathrm{GHZ}_j}$.
    
    \item[\textbf{Step 2:}] Setting up distributed angle ($\alpha$).

     This is one of the most important steps towards the final execution of the QAN. For each GHZ state $j$, a global angle $\alpha_j$ is selected and split into secret shares using quantum secret sharing. Each party $i$ receives a sub-share $\alpha_j^i$ such that,
    \begin{equation}
        \sum_{i=1}^n \alpha_j^i \equiv \alpha_j \mod 2\pi.
    \end{equation}
    This ensures that no single party learns the complete value of any $\alpha_j$.

    \item[\textbf{Step 3:}] Applying rotation operator to the qubit. 

    Each party $i$ applies a local rotation $R_z(\alpha_j^i)$ to their qubit $q_j^i$ in every GHZ state $j \in \{1, 2, \dots, n\}$.
    \begin{equation}
        R_z(\alpha_j^i) = e^{-i\alpha_j^i Z/2}.
    \end{equation}
    However, if a party is the \textit{notifier} (e.g., Alice) and wishes to notify an anonymous receiver associated with GHZ state index $r$, then for $j = r$ she modifies her rotation,
    \begin{equation}
        R_z(\alpha_r^i + \delta), \quad \text{with probability } P_z,
        \label{rotation}
    \end{equation}
    This phase shift, $\delta$, allows for a change in the parity for the receiver ($r$) to detect a notification, while still preserving the anonymity of the notifier. 

    \item[\textbf{Step 4:}] Applying Hadamard gate and performing measurement. 
    After applying the Hadamard $H$ to each qubit $q_j^i$ and measuring in the computational basis to obtain bits
    \begin{equation}
        \{m_j^i \mid j=1,\dots,n\}
    \end{equation}
    user $i$ takes all $n$ bits (including $m_i^i$) and applies a private random permutation $\pi_i$ before broadcasting.  In other words, they broadcast
    \begin{equation}
        \{\,m_{\pi_i(1)}^i,\;m_{\pi_i(2)}^i,\;\dots,\;m_{\pi_i(n)}^i\}
    \end{equation}
    so that no one--neither other users nor eavesdroppers--can tell which bit was the self-measurement bit for any particular user.

    \item[\textbf{Step 5:}] Calculating post-measurement parity.
    Once every user has broadcast $n$ permuted bits, each node collects the full multi-set of $n^2$ bits for GHZ$_j$.  They then compute the global parity
    \begin{equation}
         m_j = \bigoplus_{i=1}^n m_j^i,
    \end{equation}
    summing over all $i$ (including the self-measurement bit).  This ensures that the parity truly reflects all $n$ measurement outcomes, so a flip in round $j$ signals a notification for user $j$.
    
    \item[\textbf{Step 6:}] Repetition of the method. 

    Each party checks the parity value $m_j$ for each GHZ state. The intended receiver (Bob) would check his GHZ index and calculate $m_r$. If $m_r=1$ in any round, Bob would conclude that he got a notification. 
 
\end{enumerate}

Figure. (\ref{fig:qan-diag}) presents a schematic representation of the working of the modified QAN protocol. The QAN protocol can broadly be divided into four stages as represented in the Figure. (\ref{fig:qan-diag}). In the previous approach of the QAN protocol (as in  \cite{khan2020quantum}), all the users in the network know the qubit index assigned to each user. However, distributing $n$, $n$-partite GHZ states with known qubit assignments compromises overall anonymity. To address this, we modify the approach such that each user distributes their own set of $n$ $n$-partite GHZ states for which the user assignments are only known to the distributor. We show that this preserves the anonymity of both the sender and the receiver, even if we consider the network consisting of semi-honest parties.

\begin{enumerate}
    \item Each user in the network shares one $ n$-partite GHZ state, and for the GHZ state shared by user $U_j$, they know the indexing for each qubit and user in the network (as explained in the next section). 
    \item The secret share of the angle is only done once; this shall be used throughout all the GHZ states. 
    \item Now, if Alice wants to notify Bob in the network, she would use the $n$ $n-$partite GHZ state shared by her, such that she knows which qubit corresponds to Bob. Every other user would operate on all the $n$ $n-$partite GHZ states they own, except the one that they shared. 
    \item Everyone would measure all of these GHZ states, and finalize the QAN protocol. 
    \item Upon measurement and final XOR calculation, Bob would find a notification from the state Alice shared, and nothing from the other states. 
\end{enumerate}

\subsection{Example Case: 3-User Network}
We consider a basic network consisting of $3$ users: Alice, Bob, and Charlie. The exact steps would be, 
\begin{enumerate}
    \item Each of them prepares and distributes a $3-$partite GHZ state, 
        \begin{equation}
            |\text{GHZ}\rangle = \frac{1}{\sqrt{2}}(\ket{000} + \ket{111})
        \end{equation}
        \begin{itemize}
            \item Alice distributes GHZ\textsubscript{A}, Charlie gets one qubit, Bob gets one, and she keeps one herself.
            \item Bob distributes GHZ\textsubscript{B}, sending qubits to Alice and Charlie.
            \item Charlie distributes GHZ\textsubscript{C}, similarly sharing qubits.
        \end{itemize}
        Each of these users knows the qubit mapping for the GHZ state they shared. So, for $GHZ_A$ \textit{only Alice} knows the mapping, 
        \begin{equation}
            \text{index 0} \rightarrow \text{Alice},\quad
        \text{index 1} \rightarrow \text{Bob},\quad
        \text{index 2} \rightarrow \text{Charlie}
        \end{equation}

\item If \textit{Alice} wants to inform \textit{Bob}: she uses GHZ$_A$, i.e., the GHZ state she prepared, and applies the modified rotation $R_z(\alpha_j+\delta)$ with probability $P_z$ for the index corresponding to Bob. For the other two GHZ states, GHZ$_B$ and GHZ$_C$, Alice applies just the normal rotation ($R_z(\alpha_j)$). 

\item Bob and Charlie perform similar rotation ($R_z(\alpha_j)$) except the one they created, i.e., Bob does not do anything with GHZ$_B$, and Charlie does not do anything with GHZ$_C$.

\item Hadamard operation and measurement.  Each party does the following,

\begin{itemize}
    \item Applies Hadamard $H$ to all GHZ qubits they own.
    \item Measures them on a computational basis.
    \item Publicly shares measurement outcomes.
\end{itemize}

\item For each GHZ state, the parity is computed:
\begin{equation}
     m_j = m_j^A \oplus m_j^B \oplus m_j^C
\end{equation}
where $m_j^X$ is the measurement bit from user $X$ on GHZ\textsubscript{j}.
\item Bob gets the notification. Bob measures the parity from all the GHZ states he received, i.e., from Alice and Charlie. He should see a notification from one of the GHZ states that he received. 
\end{enumerate}
This example can be extended to $n$-user network, and the generalized protocol as presented in \textit{Algorithm \ref{alg:QAN_comp}}.

\begin{algorithm}[ht!]
\caption{Modified $n$-user Distributed QAN}
\label{alg:QAN_corrected}
\begin{algorithmic}[]
\Procedure{QAN\_Protocol}{$n,\;u,\;r,\;\{\ket{\mathrm{GHZ}_j}\}_{j=1}^n,\;\{\alpha_j^{(i)}\}_{i=1}^n,\;P_z$}
  \Comment{$u$ is the notifier’s user index, $r$ is the GHZ index to notify.}

  \For{each user $i=1,\dots,n$}                        \Comment{Rotation phase}
    \For{each GHZ index $j=1,\dots,n$}
      \If{$i = j$} 
        \State $\textbf{continue}$                   \Comment{Skip your own GHZ}
      \ElsIf{$i = u$ and $j = r$}                  \Comment{Notifier injecting notification}
        \State With probability $P_z$: apply
               $R_z\bigl(\alpha_j^{(i)}+\pi\bigr)$
        \State With probability $1-P_z$: apply
               $R_z\bigl(\alpha_j^{(i)}\bigr)$
      \Else                                             \Comment{All other cases}
        \State Apply $R_z\bigl(\alpha_j^{(i)}\bigr)$
      \EndIf
    \EndFor
  \EndFor

  \Statex
  \Comment{Hadamard, measurement, and broadcast}
  \For{each user $i=1,\dots,n$}
    \State Apply Hadamard $H$ to each qubit $q_j^i$
    \State Measure to get bits $\{m_j^{(i)}\}_{j=1}^n$
    \State Broadcast all $m_j^{(i)}$ for which $j \neq i$
  \EndFor

  \Statex
  \Comment{Parity computation and local notification check}
  \For{each GHZ index $j=1,\dots,n$}
    \State Compute $m_j = \bigoplus_{i=1}^n m_j^{(i)}$
    \State Let $v=\Phi_j(r)$ be the receiver slot for GHZ$_j$
    \If{$m_j = 1$ and current user $i = v$}
      \State User $v$ concludes “Notification received.”
    \EndIf
  \EndFor
\EndProcedure
\end{algorithmic}
\label{alg:QAN_comp}
\end{algorithm}

\section{Results}
{In this section, we present the simulation results based on Algorithm} \ref{alg:QAN_comp}, {with specific parameters presented in the Table.} \ref{tab:params} {along with the noise models used, and some possible attacks on the modified QAN system.} {The values $P_z = 0.1, 0.3, 0.45$ were selected to represent low, medium, and high notification actuation probabilities, respectively, enabling evaluation of protocol behavior across different operating regimes.} {Before presenting the simulation outcomes, we briefly discuss the theoretical basis underlying the expected results. The notification detection process in the QAN protocol depends on the interference of shared GHZ states, where local phase rotations determine whether the receiver observes a parity flip. Channel noise, modeled through dephasing and depolarizing effects, reduces the coherence of these GHZ states and thereby lowers the contrast between notification and no-notification cases. The performance of GHZ-based communication schemes is strongly influenced by the fidelity and coherence of the entangled states used, as discussed in earlier works on noisy multipartite entanglement and quantum communication}  \cite{guo2023noise} {As coherence decreases, the overall probability of successful notification detection is expected to drop, while false-positive events become more likely. The following section presents simulation results that illustrate these effects and highlight the improved robustness of the modified QAN protocol compared to a prior approach.}

\subsection{Simulation Results}
\label{Sec:sim}

\begin{figure*}[h!]
    \centering
    \includegraphics[width=0.7\linewidth]{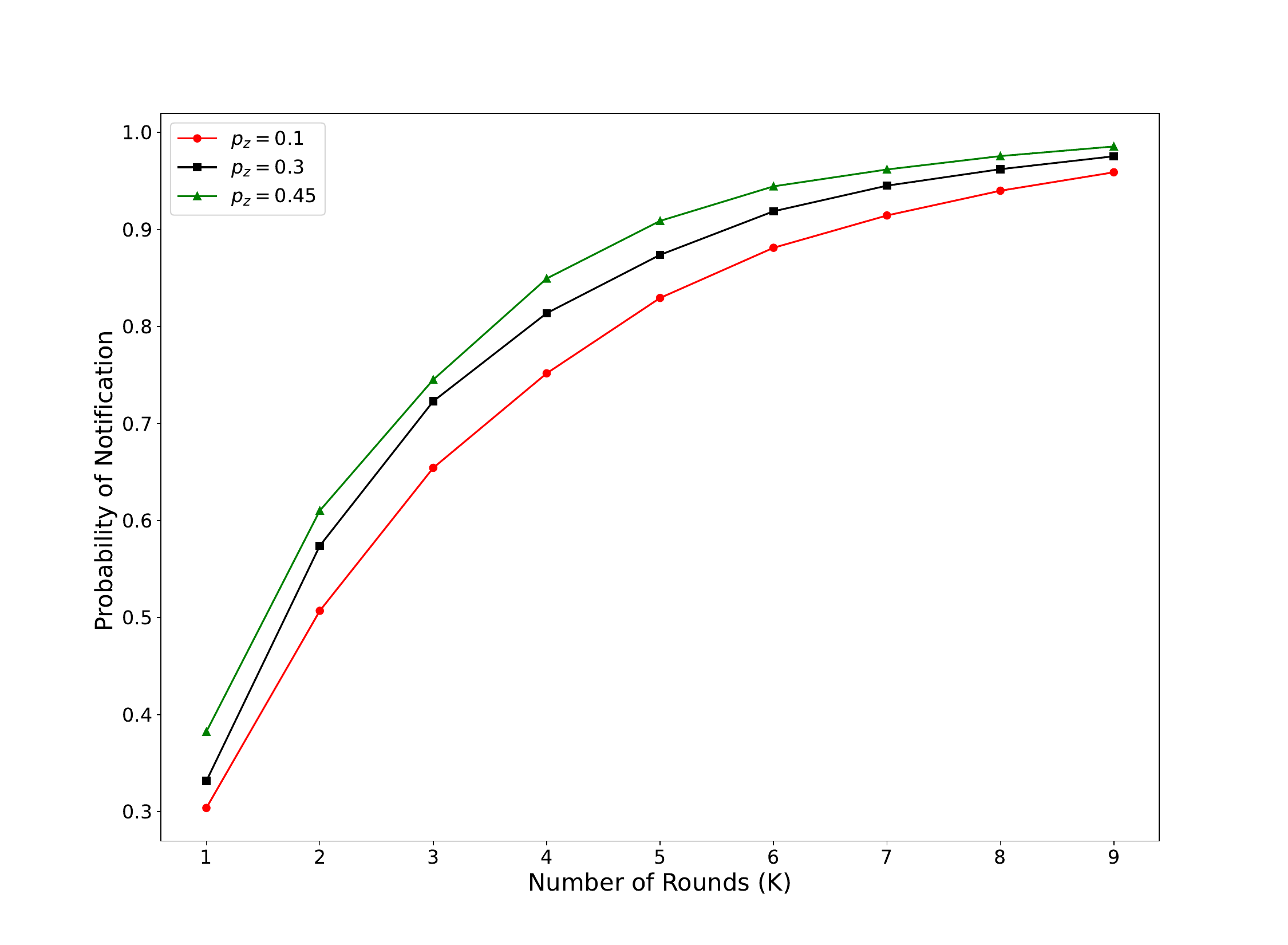}
    \caption{Simulation results for QAN protocol outlined in Sec[\ref{Sec:qan}] for detection probability of notification received for increasing number of runs for different values of $P_z$, i.e., probability of application of the rotation operator by the notifier for the receiver index. }
    \label{fig:qan}
\end{figure*}

\begin{figure*}[h!]
    \centering
    \includegraphics[width=0.6\linewidth]{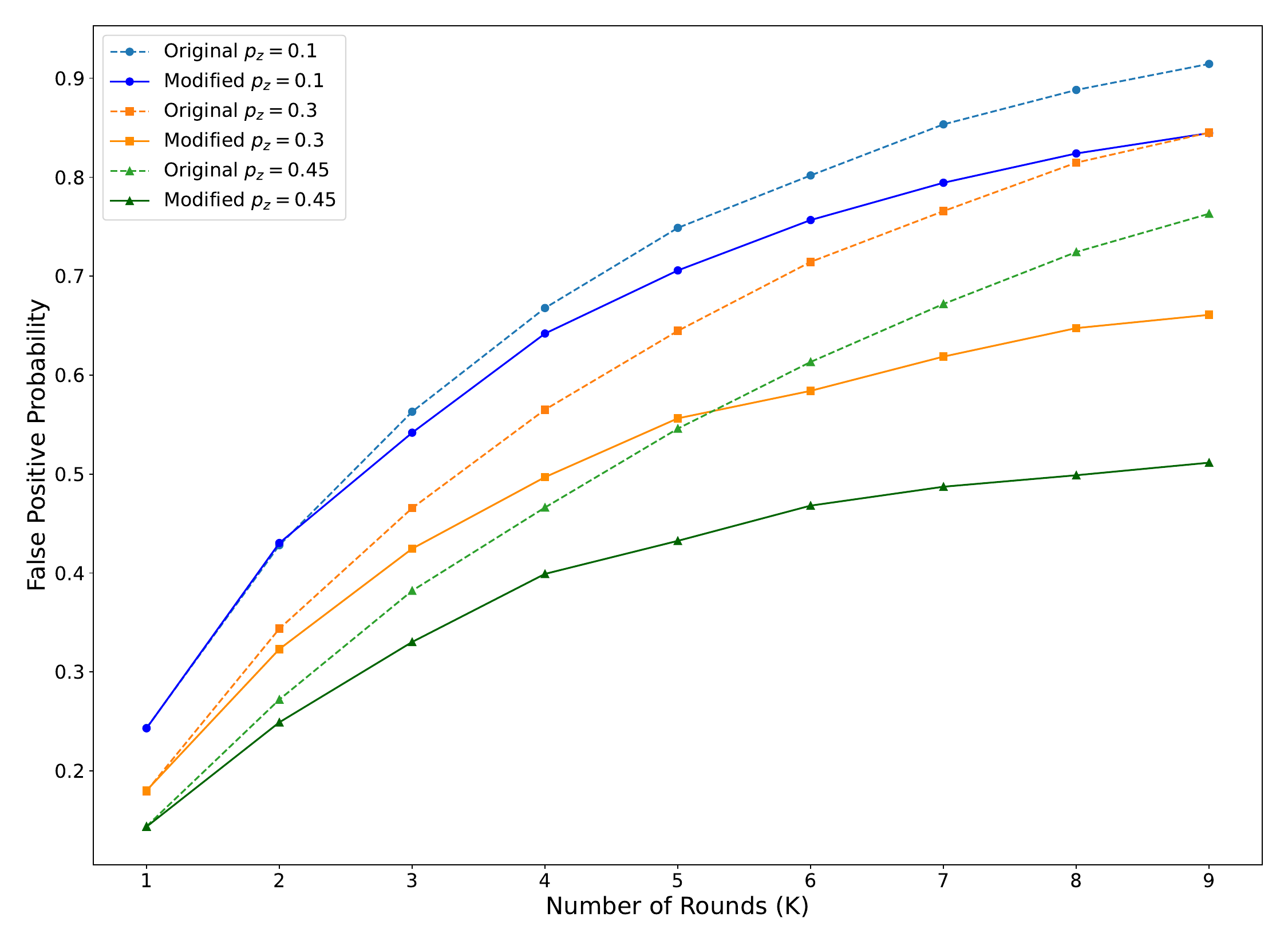}
    \caption{This figure shows the false positive rate of notification detection in the presence of noise models mentioned above for the modified QAN protocol proposed in this work, compared to the QAN presented by Khan et al.  \cite{khan2020quantum}. The false positive rate is calculated as any notification detected by an unintended party, as a result of different noise models present in the network.}
    \label{fig:compare}
\end{figure*}

 For simulation, we focus on the GHZ state shared by the notifier, Alice, in an $n=4$ network. Through these simulations, we present the performance of our proposed QAN protocol, and the comparison of the modified QAN protocol to the protocol presented in  \cite{khan2020quantum}. Through the simulations, we present the performance of the modified QAN protocol (in Figure. (\ref{fig:qan})), and the performance comparison of the modified QAN protocol with respect to the original QAN protocol under depolarizing noise models (in Figure. (\ref{fig:compare})). The other simulation parameters are as follows,

\begin{table}[h!]
\centering
\caption{Simulation parameters}
\label{tab:params}
\renewcommand{\arraystretch}{1.4} 
\setlength{\tabcolsep}{12pt}      
\normalsize                        
\begin{tabular}{|l|c|p{9.5cm}|}
\hline
\textbf{Parameter } & \textbf{Value(s)} & \textbf{Notes} \\
\hline
Secret angle shares ($\alpha_i$) & $0$ $\forall$ $i$ &
No net pre-rotation on the initial state; sets baseline phase to zero. \\
\hline
Number of users ($n$) & $4$ &
Network of four parties. \\
\hline
Notification phase kick ($\delta$) & $\pi$ &
Additional probabilistic rotation applied at the receiver index ($j=r$). \\
\hline
Rounds per experiment ($K$) & $[1,9]$ &
Used to observe how detection probability accumulates with repetitions. \\
\hline
Notifier actuation probability ($P_z$) & $[0.1,\ 0.3,\ 0.45]$ &
Probability that the notifier applies $\alpha'=\alpha+\delta$ on the receiver’s index in a round. \\
\hline
\end{tabular}
\end{table}

Figure. (\ref{fig:qan}) shows the curves for $3$ different values of $P_z$ for increasing number of $K$. In this figure, we have plotted the probability of a notification detected by the intended receiver when the sender applied the rotation for receiver index qubit with a probability of $P_Z$, plotted against the number of repetitions of QAN rounds, $K$.

\begin{figure}[h!]
    \centering
    \includegraphics[width=0.65\linewidth]{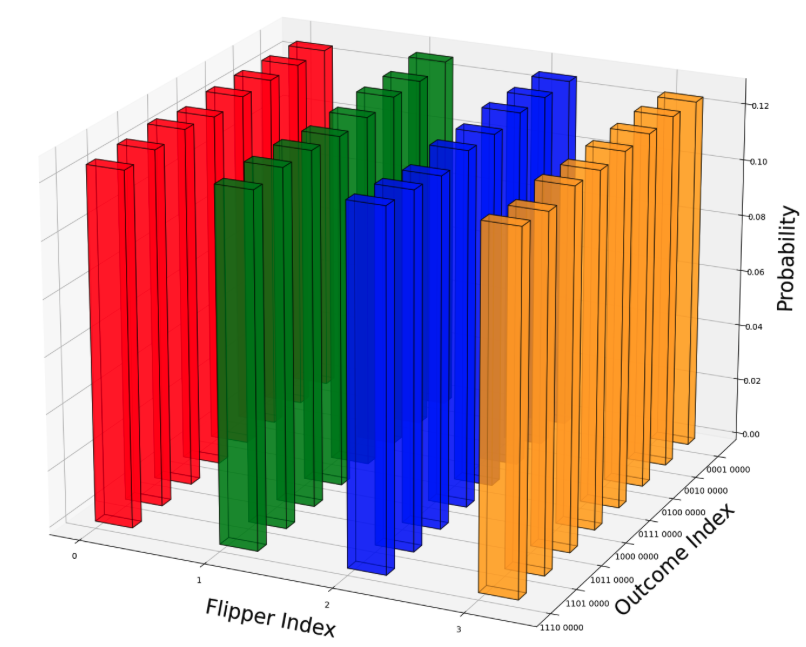}
    \caption{The probability distribution of different states in the QAN protocol shows a uniform distribution. It, thus, preserves the anonymity of the notifier. Here, we plotted the probability distribution of various strings averaged over $1000$ simulation runs.}
    \label{fig:anon}
\end{figure}

From the Figure. (\ref{fig:qan}), We notice that for the increased value of the probability of rotation application, the detection of notification by the receiving party improves significantly. We notice that we can find a middle-ground between increasing $N$ values and $P_z$ values to optimize the overall notification detection probability. Figure (\ref{fig:anon}) illustrates that the QAN protocol preserves user anonymity. The $x-$axis shows the flipper index, i.e., the qubit index for possible notifier or receiver. The $y-$axis shows the outcome index, which shows the combined bit-string obtained by measuring the GHZ state, and the $z-$axis presents the probability of different outcome strings when measured by different flipper indices (i.e., users). The protocol achieves a high notification detection rate across various $P_Z$ and $K$ values while maintaining notifier anonymity, as shown in the figures mentioned above.

\subsection{Noise Integration}
\label{Sec:noise}

To study the extent of the effectiveness of the protocol, we introduce depolarizing noise models and present the comparison of the false-positive notification detection in our modified QAN approach (using rotation operators) vs the QAN protocol presented in  \cite{khan2020quantum}. The dephasing noise model is explored because it sets the detection gap and therefore how many rounds are needed for reliable detection on real links. {Depolarizing noise is also an appropriate model for the quantum components considered in this work, as it captures the aggregate effects of imperfect single and two-qubit gate operations, control inaccuracies, and residual channel interference commonly observed in near-term quantum hardware.} For this simulation, we used the following noise model,

\begin{enumerate}
    \item \textit{Single qubit Depolarizing noise}:

    For a single qubit gate, $G\in \{H, U, Z  \}$, ($U$ is a possible unitary matrix operator, such as a rotation operator) is applied as,
    \begin{equation}
    \small
    \mathcal{E}_{\text{single}}(\rho) = (1 - p)\rho + \frac{p}{3} \left( X\rho X^\dagger + Y\rho Y^\dagger + Z\rho Z^\dagger \right),
    \label{eq:depolar}
    \end{equation}
    where,
    \begin{itemize}
        \item $\rho$ is the density matrix of the quantum state
        \item $p$ is the probability of depolarizing noise channel being applied (for simulation we used $p=0.01$)
        \item $X,Y,Z$ are Pauli's matrices
    \end{itemize}

    \item \textit{Two-qubit Depolarizing noise}: 

    For the controlled-X (CX) gate, a two-qubit depolarizing channel is applied as, 
    \begin{equation}
        \small
        \mathcal{E}_{\text{two}}(\rho) = (1 - p)\rho + \frac{p}{15} \sum_{i=1}^{15} P_i \rho P_i^\dagger,
        \label{depol-2}
    \end{equation}
    where, 
    \begin{itemize}
        \item $\rho$ is the density matrix of the quantum state
        \item $p$ is the probability of depolarizing noise channel being applied (for simulation we used $p=0.02$)
        \item $P_i$ are the $15$ non-identity two-qubit Pauli operators ($X\otimes Y$, $Z\otimes X$, $X\otimes I$, etc.)
    \end{itemize}

\end{enumerate}

{While the simulation results focus on a network of $n = 4$ users, the proposed QAN protocol naturally generalizes to larger networks. The quantum resource overhead scales as $\mathcal{O}(n^2)$, since each of the $n$ users distributes one $n$-partite GHZ state. The communication complexity scales quadratically, as each user broadcasts measurement outcomes associated with all received GHZ states.}

\subsection{Attacks}
The anonymity of the users participating in the QAN protocol mentioned in the above section can be analyzed for some attacks where we consider semi-honest participants and compromised users. Here, we show that even if we consider all the users in the network to be semi-honest, the anonymity of any sender and receiver is preserved.

\subsubsection{Assuming Semi-Honest Parties}
We assume the set of users to be $u = \{ 1,2,3, \dots, n\}$. For each user $j$, we consider a $n-$partite GHZ state represented as follows,
\begin{equation}
    \ket{\text{GHZ}_j} = \frac{1}{\sqrt{2}}\left(\ket{0}^{\otimes n}+ \ket{1}^{\otimes n} \right),
\end{equation}
which is prepared and distributed by user $j$. The user $j$ has an associated secret mapping $\Phi_j$ which lets them know which qubit belongs to which user in the system. This can be written as, 
\begin{equation}
    \Phi_j: \{1,2,\dots, n\}\to u.
\end{equation}
This shows the mapping (bijective mapping) of each qubit in $\ket{GHZ_j}$ belongs to one user. This mapping, as mentioned, is only known to the distributor, $j$.

Now, we consider a general notifier, $u$, and a general receiver, $v$. The user $u$ will use $\ket{GHZ_u}$, i.e., the state they distributed. For the secret mapping, $\Phi_u$, there exists a unique index, $r$, such that, 
\begin{equation}
    \Phi_u(r) = v. 
\end{equation}
During the QAN stage, every party modifies its qubit by applying a rotation operator ($R_z(\alpha_j$)), where $\alpha_j$ is the secret share of angle distributed to each user in the network. However, the notifier $u$ modifies the rotation on her qubit in position $r$ by applying a rotation of ($R_z(\alpha_j+\delta)$) with probability of $P_z$. All other parties, upon receiving their qubits from $\ket{GHZ_u}$ apply the standard $R_z(\alpha_j$). 

After the Hadamard gate application, and measurement is done (in computational basis), all parties declare the result for $i^{th}$ qubit, i.e., every other qubit except their index. Let $m^i$ be the measurement outcome for $i^{th}$ qubit, so the global parity is defined as, 
\begin{equation}
     m = \oplus_{i=1}^n m^i.
\end{equation}
In an ideal no-noise scenario, for the no-notification case $m=0$, with a high probability. However, for a case when notifier $u$ wants to notify $v$, the $m$ changes to $1$ with some probability. Crucial thing to note here is that only the receiver $v$ can identify the parity change for the $j^{th}$ qubit associated with them. Even if we consider semi-honest users, i.e., they try to gain information from publicly shared data, they are not able to identify any local parity change-- which only $v$ can calculate. 

\subsubsection{Compromised User}
If we consider a network with $m$ compromised users, such that $m\leq n/2$. We consider one of the compromised users to be $\mathcal{C}$. A similar argument can be used to show that the anonymity of the sender and receiver is preserved, unless the compromised party is the sender. However, even in this case, where $\mathcal{C}$ is the sender, the only information she can gain by publication announcement is the index of the receiver. However, since for each user there exists a different secret mapping defined by them, no further information can be gained by outsiders, even if half of the users in the system are compromised. 

However, if we consider a system with a large number of compromised users, it can cause some issues with the overall working of the system. Since, QAN step comes before transmission of an actual quantum-encrypted message, a compromised node can apply random rotations to random indices to disrupt the overall parity of the system. Thus, this particular disruption needs to be taken care of in the long run for practical deployments.

 In practical hardware implementations, the fidelity of shared GHZ states may not remain ideal due to gate errors, loss, and decoherence. A small drop in the initial fidelity would slightly lower the contrast between the notification and no-notification cases. This means the detection probability decreases marginally, while the false-positive rate shows a small increase. Although a detailed quantitative analysis is beyond the current scope, this trend follows the expected behavior of multipartite GHZ states under noise-heavy conditions and emphasizes the importance of maintaining high state-preparation fidelity for reliable network performance.

\section{Discussions}
\label{Sec:Conclusion}
The modified QAN protocol has been presented in this work, and the performance has been validated through several experiments, and security under several plausible attacks has also been studied. Now, we will discuss how this QAN protocol can be integrated with the QuANet framework, such that we can achieve some protection from compromised switches in a network by achieving switch-independence. 

\subsection{Integrating QAN with the QuANet Framework}
\label{Sec:integrate}

\begin{figure*}[h!]
    \centering
    \includegraphics[width=0.6\linewidth]{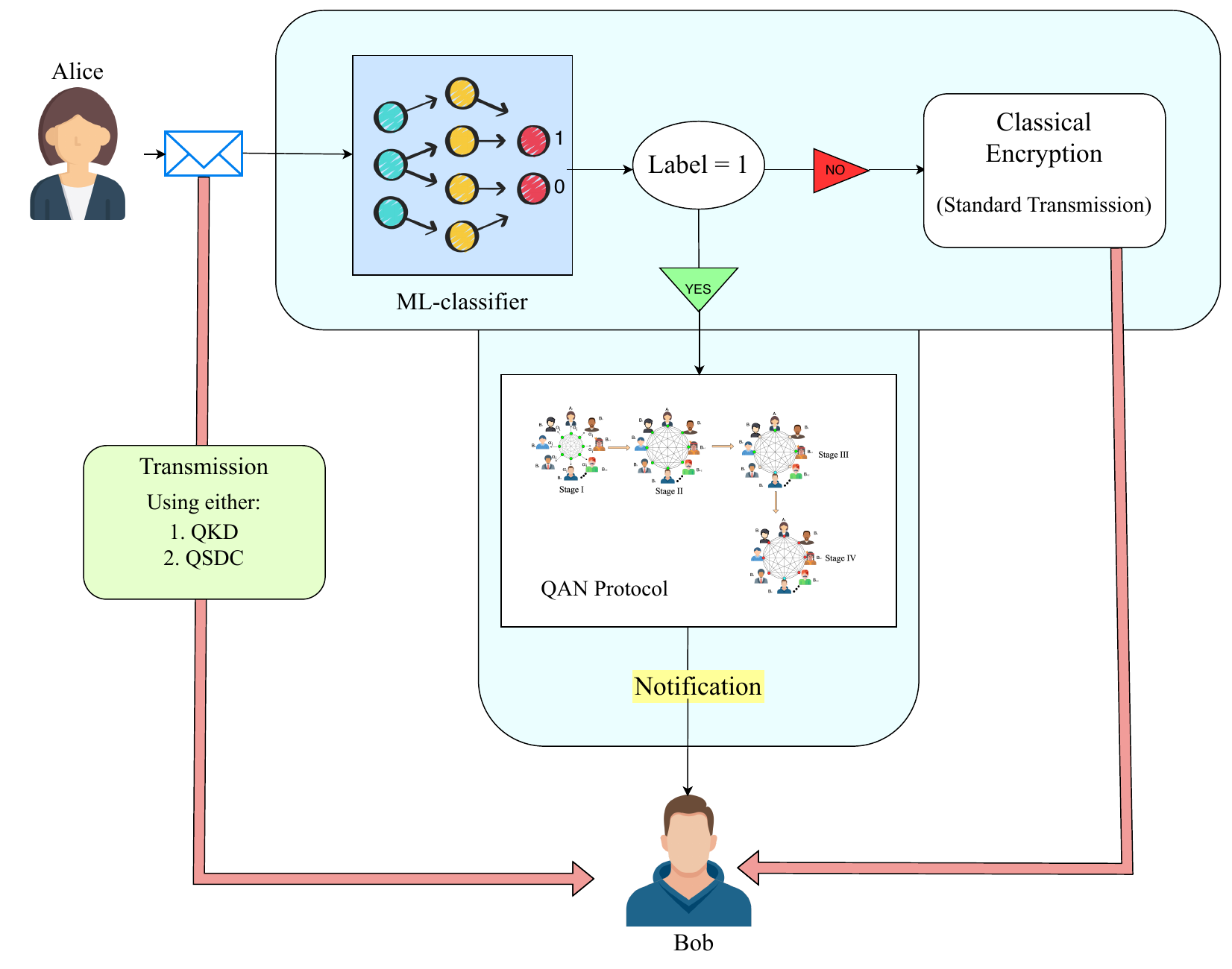}
    \caption{A schematic diagram of the combined protocol that integrates the QAN protocol into the idea of quantum augmented network presented in  \cite{jha2024ml}. The diagram is divided into two parts: (1) the part outlined in \textit{blue box} outlines the protocol that's executed before actual transmission, (2) the red arrows mark the transmission and encryption method. In the first half of the protocol, Alice gets a privacy label for her message, and based on that, she either initiates the QAN protocol or she uses classical encryption techniques (for example, AES, etc.) to transmit her message to Bob. Once QAN is performed (for private messages), we move on to the transmission phase. In this phase, if QAN was used, Alice and Bob either start a QKD run or Alice sends a packet with a quantum payload to Bob right after the notification.}
    \label{fig:qancombined}
\end{figure*}

The integration of the QAN protocol with the quantum-augmented network (QuANet) protocol helps us address some of the key security weaknesses of the QuANet structure.  The QuANet framework presented in  \cite{jha2025towards} uses machine-learning classifiers to assign a privacy label to each outgoing communication. If a communication contains some private content, then quantum encryption is used to transmit it securely over the network. This quantum-encrypted payload is sent in a packet with quantum and classical headers as outlined in  \cite{jha2024ml}. However, having such packets flagged as ``quantum encrypted" can have security concerns if switches are compromised.  A high-level schematic diagram for the protocol is also presented as a Figure. (\ref{fig:qancombined}) The three main steps of this integration are as follows:
\begin{enumerate}
    \item ML-Based Privacy classifier: determines if a message contains any private content. The definition of ``privacy" is user dependent, for example, personal identifiers, medical and financial identifiers, etc.
    \item Quantum Anonymous Notification Protocol: If any message is classified to have private content, this should be triggered by the sender (referred to as Alice across this work). The receiver (Bob) should receive a notification. 
    \item The receiver (Bob) informs the switch to bypass incoming packets directly to the quantum gateway, where the quantum payload can be processed. This avoids the switch to interact with the packet, such as interacting with the packet header. 
\end{enumerate}

The integration of the QAN protocol with the QuANet framework is formally presented in \textit{Algorithm \ref{alg:Quanet}}. The protocol begins with an ML-based decision procedure that evaluates the privacy level of a composed message M. If the message is classified as non-private (L=0), classical encryption is applied before transmission. Otherwise, if the message is deemed private (L=1), the Quantum Anonymous Notification (QAN) protocol is invoked.

\begin{algorithm}[h!]
\caption{Integrated Protocol}
\begin{algorithmic}
\Procedure{ML-Based Decision}{}
    \State Compose message $M$.
    \State $L \gets \text{ML}(M)$ \Comment{$L=0$: non-private; $L=1$: private.}
    \If{$L=0$}
        \State Apply classical encryption to $M$.
        \State Transmit $M$.
        \State \Return.
    \Else
        \State \textbf{Proceed to QAN protocol.}
    \EndIf
\EndProcedure

\Procedure{Quantum Anonymous Notification}{}
    \State \textbf{Assumption:} GHZ states are pre-distributed.
    \State Alice triggers QAN to notify Bob.
    \State QAN is executed by the network.

\EndProcedure
\Procedure{Switch Bypass Activation}{}
    \State Route subsequent packets directly to the quantum gateway.
\EndProcedure
\Procedure{Packet Arrival}{}
\State Packet with quantum payload arrives at the quantum gateway
\State Decryption and Measurement Done
\EndProcedure
\end{algorithmic}
\label{alg:Quanet}
\end{algorithm}

By using this integrated approach to a quantum augmented network, the protocol achieves \textit{switch independence}, which is particularly beneficial in the case where there are compromised switches in the system. This also allows us to remove any information from the packet headers that identifies any packets with private contents, i.e., quantum-encrypted payload. This would help us reduce any directed attacks that can be performed by the adversaries as a result of these compromised switches in the network.

\subsection{Conclusion}
In this work, we presented an improved Quantum Anonymous Notification (QAN) protocol that can be easily integrated with a Quantum-Augmented Network (QuANet) framework. The proposed QAN protocol uses quantum secret sharing to establish a secret share of the angle that every user in the system holds, which is then modified by any sender to anonymously notify the receiver using pre-shared GHZ states. The integration of QAN with the QuANet framework allows us to modify the pre-proposed structure of message packets, such as headers, payload, etc. The header information from these packets can be exploited by compromised switches to gain some contextual information, as this framework uses selective quantum encryption for messages containing private information. The adversaries can thus use such information to either delay or drop these packets at these compromised switches. Use of QAN before actual transmission would allow the receivers to bypass incoming messages from the switches, and also help in reducing the information that is provided in the headers. This is what we like to refer to as \textit{switch independence}.

The simulation results as shown in the Figure. (\ref{fig:qan}) and Figure. (\ref{fig:anon}) shows that the QAN protocol achieves decent notification-detection probability even for smaller rounds ($K$). We also present a comparison of the working of our QAN protocol to one of the earlier defined approaches under a couple of noise models. From the Figure. (\ref{fig:compare}), we can see that the false-positive rate for notification detection is significantly different (about $10\%$ for any $K\geq3$) for our approach vs the older one. We also present an analysis for two possible attack scenarios for our QAN protocol, and show that the anonymity of the sender-receiver pair is maintained even in the case of all users being semi-honest, or users in the system being compromised. Future works can focus on active attacks, such as rotation poisoning or targeted $Z-$gate attacks. For example, we consider the case of the last-speaker steering attack. Here, an adversary holds on till they can see others' broadcasts and then chooses their bit to cause the final parity changes. These attacks can be further avoided by providing each party with certifications and signatures to avoid this class of attacks. {If we consider the case of a rotation poisoning attack by a single compromised node that slightly perturbs its local rotation each round, we notice that the shared GHZ coherence degrades, false notify events become more likely, true notifications are harder to confirm, and convergence slows as the network size grows. A simple mitigation is to randomize local bases across rounds, watch for sudden drops in measured coherence, and aggregate a few rounds by majority vote so one misbehaving node cannot drive the outcome.}

These modification to the QuANet architecture not only increases the overall security of the system, but also provide us with a scalable and more robust framework than the stand-alone QuANet framework. The future work can look at network simulation for this work to gain more insights about possible bottlenecks and shortcomings in a fully-fledged network scenario. There are, however, some limitations of this work that needs to be further studied in upcoming studies such as the analysis of fidelity of sharing GHZ states under noise-heavy hardware, a more thorough analysis of the performance of switch-independence in different network configurations, and thus, development of any further protocols which can streamline the integration of our QAN protocols with other quantum secure direct communication (QSDC) protocols necessary for quantum secure communication \cite{sun2025quantum, zou2014three}. {In addition to this, the modified QAN can be used in the development of quantum augmented microgrid structures, where we integrate the use of quantum components such as quantum key distribution, quantum anonymous notification, and quantum random number generation to develop a secure quantum-augmented microgrid. }

\section*{Authors Contributions}
N.J. is the primary author of the manuscript and received intellectual inputs from A.P. and M.S. to guide the research. All authors edited the manuscript.

\section*{Data availability statement}
All data generated or analysed during this study are included in this published article [and its supplementary information files].

\section*{Additional Information (Competing Interests)}
None

\nocite{*}
\bibliography{main}
 
\end{document}